\documentclass[prl,preprint,superscriptaddress]{revtex4}
\usepackage{graphicx}
\usepackage{dcolumn}
\usepackage{amssymb}
\usepackage{bm}
\usepackage{pifont}
\usepackage{epsfig}
\usepackage{psfrag}
\usepackage{epstopdf}
\usepackage{wasysym}
\usepackage[usenames]{color}
\usepackage[symbol]{footmisc}
\def\correspondingauthor{\footnote{rajeshg@jncasr.ac.in.}}

\begin{document}

\title{Unraveling the Role of Frictional Contacts and Particle Orientational Order During Shear-thickening in Suspensions of Colloidal Rods}

\author{Vikram Rathee}
\affiliation{Department of Physics, Indian Institute of Science, Bangalore - 560012, INDIA}
\affiliation{Department of Physics, Georgetown University, Washington D.C.20057, U.S.A.}
\affiliation{ Institute for Soft Matter Synthesis and Metrology, Georgetown University, Washington D.C.20057, U.S.A.}
\author{Srishti Arora}
\affiliation{International Centre for Materials Science, Jawaharlal Nehru Centre for Advanced Scientific Research, Jakkur, Bangalore - 560012, INDIA}
\author{Daniel L.  Blair}
\affiliation{Department of Physics, Georgetown University, Washington D.C.20057, U.S.A.}
\affiliation{ Institute for Soft Matter Synthesis and Metrology, Georgetown University, Washington D.C.20057, U.S.A.}
\author{Jeff S. Urbach}
\affiliation{Department of Physics, Georgetown University, Washington D.C.20057, U.S.A.}
\affiliation{ Institute for Soft Matter Synthesis and Metrology, Georgetown University, Washington D.C.20057, U.S.A.}
\author{A. K. Sood}
\affiliation{Department of Physics, Indian Institute of Science, Bangalore - 560012, INDIA}
\affiliation{International Centre for Materials Science, Jawaharlal Nehru Centre for Advanced Scientific Research, Jakkur, Bangalore - 560064, INDIA}
\author{Rajesh Ganapathy\correspondingauthor{}}
\affiliation{International Centre for Materials Science, Jawaharlal Nehru Centre for Advanced Scientific Research, Jakkur, Bangalore - 560064, INDIA}
\affiliation{School of Advanced Materials (SAMat), Jawaharlal Nehru Centre for Advanced Scientific Research, Jakkur, Bangalore - 560064, INDIA}
\date{\today}

\draft

\date{\today}

\begin{abstract}
There is now convincing evidence that inter-particle frictional contacts are essential for observing shear-thickening in concentrated suspensions of compact particles. While this has inspired many strategies to tailor the rheology in these systems, in the more general case, viz-\'a-viz suspensions of anisotropic particles, the mechanism of shear-thickening remains unclear. Here through simultaneous measurements of the bulk viscosity and the first Normal stress difference, we show that strong shear thickening in suspensions of colloidal rods is accompanied by large positive normal stresses, indicating the formation of a system-spanning frictional contact network. We also find that flow in the shear-thickening regime is unsteady and shows a rather rich time-dependence. By carrying out single particle-resolved confocal rheology measurements, we provide compelling evidence that this rheological chaos arises from a strong coupling between the imposed flow and particle orientational order. Building on these observations, we designed colloidal rods with temperature-tunable tribological properties and demonstrate the feasibility of achieving \textit{in-situ} control over suspension rheology. These findings show that the interplay between orientational order and frictional interactions plays a critical role in the shear thickening of dense suspensions of colloidal rods.
\end{abstract}

\maketitle
When subjected to shear stresses beyond a material-dependent threshold value, most dense suspensions shear-thicken - their viscosity $\eta$ increases with the stress $\sigma$ \cite{barnes_JR,wagner_PT,brown_RPP,brown_NatMat}. Depending on the particle volume fraction $\phi$ the increase in $\eta$ can either be gradual or abrupt. Such a response to flow not only finds extensive use in applications ranging from the design of ballistic armor \cite{wagner_JMatSci} to electrolytes for impact-resistant automobile batteries \cite{Ye_ACSApplNanoMat}, but also has detrimental effects, for instance, during pipe flow of dense slurries where it can result in clogging. Elucidating the microscopic underpinnings of shear-thickening and devising strategies to control this behavior has therefore been at the forefront of research efforts. The consensus from these studies, a substantial fraction of which are on suspension of spherical or compact particles, suggests that while the hydrocluster mechanism \cite{Brady_JFM2000,Itai_Science} can account for a nominal increase in $\eta$, a larger increase stems from particles forming a stress-induced frictional contact network \cite{Lootens_PRL,Fernandez_PRL,Denn_PRL,Cates_PRL,Mari_PNAS,Denn_JRheo,Lin_PRL,Poon_PRL2015,Hudson_PRL,vikram_PNAS}. This has resulted in a multitude of strategies where shear-thickening is passively controlled by altering particle tribological properties \cite{Hsu_PNAS} or actively tuned through complex mechanical perturbations that disrupt the formation of force chains \cite{Itai_PNAS}. Even though altering the particle shape is a powerful means to control rheological response \cite{Lettinga,Brown_PRE}, very little is known about the mechanism of shear-thickening in suspensions of anisotropic particles. An increase in the particle aspect ratio $\alpha$ reduces the $\phi$ at which shear-thickening sets in \cite{Brown_PRE}. Previous Rheo-SANS studies on suspensions of elongated particles found that particle flow-alignment was retained in the shear-thickened state and this was taken as evidence for the hydrocluster mechanism \cite{Wagner_2}. These studies, however, did not assess the first Normal stress difference $N_1$, the algebraic sign of which helps distinguish between hydrodynamic and contact contributions to the overall increase in $\eta$ \cite{Hudson_PRL}. Furthermore, scattering approaches are inherently ensemble-averaged measurements and do not capture changes in local orientational order. These concerns notwithstanding, the strong coupling between flow and particle shape anisotropy is known to result in rich time-dependent flows even in the very dilute limit and its role at higher $\phi$s and especially during shear-thickening remains unclear. Addressing this concern is of paramount importance in light of recent findings that have found unsteady flows during shear-thickening in both Brownian \cite{Lootens_PRL,vikram_PNAS} and non-Brownian suspensions \cite{Poon_JoR,Manneville} of spherical or compact particles.

Here we carried out bulk rheology and particle-resolved confocal-rheometry experiments to unravel the mechanism of shear-thickening in dense suspensions of colloidal rods \cite{KITP}. Our key results are that shear-thickening in these systems is primarily an outcome of particles forming frictional contacts. We further show that bulk flow in the shear-thickening regime is unsteady and displays strikingly complex temporal oscillations. By simultaneously imaging the dynamics of individual colloidal rods we provide indisputable evidence for the presence of strong coupling between flow and particle orientational order. Furthermore, by designing particles whose tribological properties could be altered through grafted thermo-responsive polymer brushes, we not only tune the suspension rheology \textit{in-situ} but also exploit this feature to show that shear-thickening cannot be reconciled with an order-disorder transition \cite{Hoffman}. 

\section*{Results}   
\subsection*{Shear-thickening in Colloidal Rods}
\begin{figure*}[htbp]
\centering
\includegraphics[width=0.9\textwidth]{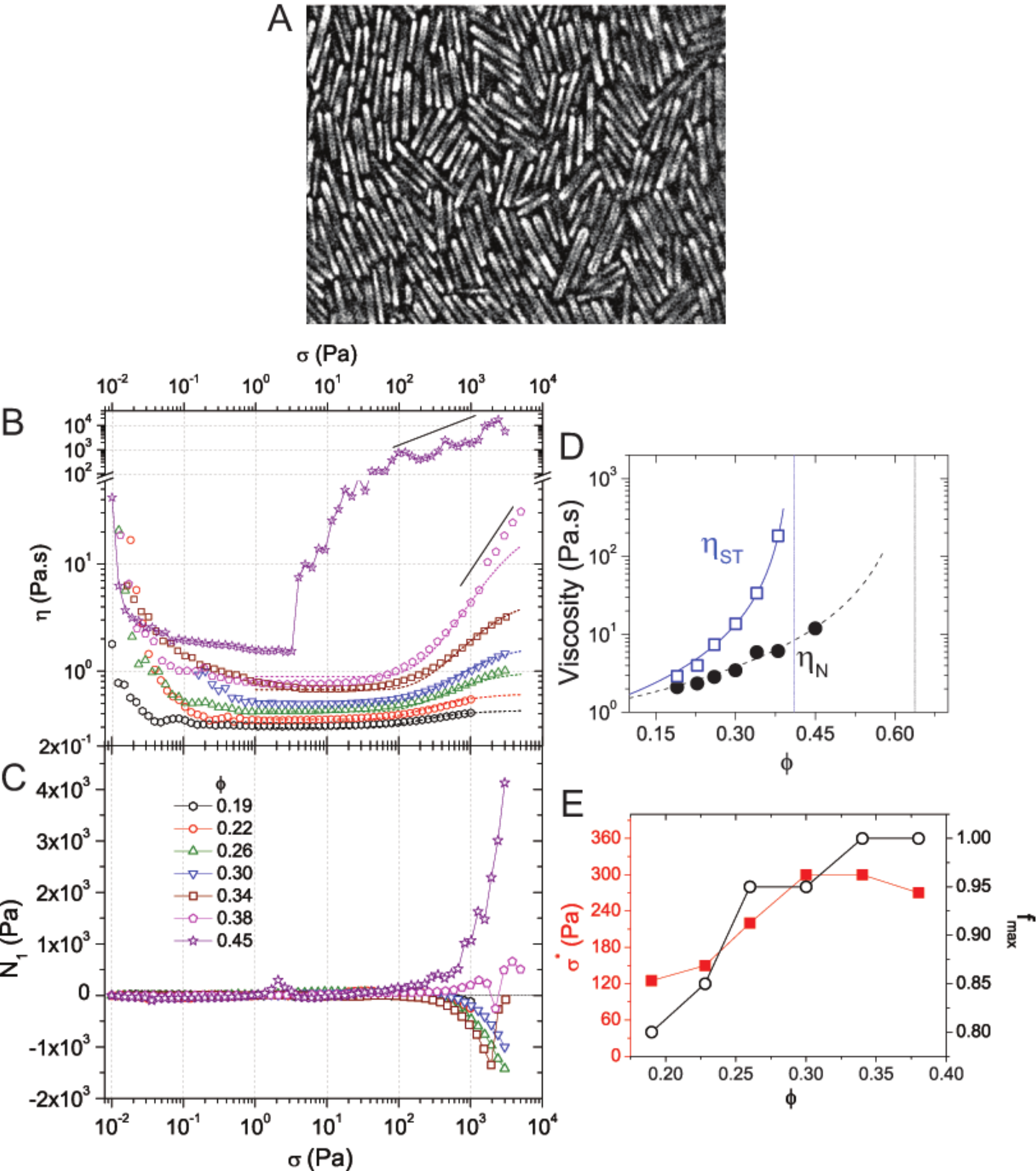}
\caption{(A) Confocal microscope image of colloidal silica rods. (B) and (C) show flow curves and the first Normal stress difference $N_1$ at various $\phi$s, respectively. The volume fractions, $\phi$, are indicated in (C). The dotted lines in (B) are fit to WC model given in eqn. \ref{eqn1} \cite{Cates_PRL}. The solid line corresponds  represent $\beta = 1$.  Please note the break in the y-axis. (D) High-shear plateau viscosity ($\eta_{N}$) and viscosity of the shear-thickened state ($\eta_{ST}$) as function of $\phi$. The dotted and solid lines show fits to $\eta_{N}$ =  (1-$\phi/\phi_{o})^{-2}$ and  $\eta_{ST}$ = (1-$\phi/\phi_{m})^{-2}$ with $\phi_{o} = 0.64$ and $\phi_{m} = 0.41$. $\eta_{ST}$ could not be determined at the highest $\phi$s where the shear-thickening plateau is not apparent. (E) $\sigma^{*}$  and $f_{max}$ versus $\phi$.} 
\label{Figure1}
\end{figure*}

We synthesized colloidal silica rods following the procedure first developed by Kuijk et al. (see Materials and Methods and Supplementary Fig. S1) \cite{Kuijk_JACS}. The rods had a typical diameter $d = 400$nm and length of $l=4\mu$m ($\alpha = l/d \sim 10$) with a size polydispersity $<10\%$ (Fig. \ref{Figure1}A). The particles were dispersed in a glycerol-water mixture (85\% glycerol by vol.) at different $\phi$s and $\eta$ was measured as a function of $\sigma$ in a cone-plate geometry (see Methods). The flow response is typical of shear-thickening particulate suspensions (Fig. \ref{Figure1}B). On increasing $\sigma$, the suspensions shear-thinned and reached a viscosity plateau, $\eta_{N}(\phi)$, which we identified with the high shear Newtonian branch. This plateau persists up to an onset stress $\sigma_c$ beyond which the suspensions shear-thickened. $\sigma_c$ is the stress required for particles to overcome repulsive interactions and come in proximity and is independent of $\phi$ as expected (Fig. \ref{Figure1}B). At even larger stresses, we encountered a second plateau that corresponds to the viscosity of the shear-thickened state, $\eta_{ST}(\phi)$. In the shear-thickening regime, $\eta\propto\sigma^{\beta}$, where $\beta$ the shear-thickening exponent is positive and steadily grows with $\phi$ but remains below unity (continuous shear-thickening) for $0.19\leq\phi<0.38$. However, for $\phi\geq0.38$, $\beta = 1$ i.e. the suspension viscosity increases at a fixed shear-rate, $\dot{\gamma}$, which is a charactisitic of discontinuous shear-thickening (DST). Remarkably, our suspensions show shear-thickening at particle loadings that are substantially smaller than those required for isotropic particles ($\beta>0$ even for $\phi = 0.19$). This observation is consistent with previous studies, which find that the larger excluded volume of anisotropic particles \cite{Philipse,Duijneveldt} in comparison to isotropic one's aid in achieving shear-thickening even at low $\phi$s.

Shear-thickening in suspensions of spherical particles is thought to arise from the formation of particle clusters mediated by hydrodynamic lubrication forces or contact friction or a combination of both. When particle clusters are stabilized by hydrodynamic forces $N_1$ is found to be negative, while $N_1$ is positive if it is due to contact friction. The behavior of $N_1(\sigma)$ at various $\phi$s is similar to suspensions of isotropic particles (Fig. \ref{Figure1}C) \cite{Hudson_PRL}. For $\phi<0.34$, N$_{1}$ hovers around zero for small applied stresses and turns negative beyond $\sigma_{c}$, a signature of dominant contribution from hydrodynamic forces to shear-thickening. However, at $\phi = 0.34$, N$_{1}$ first becomes increasingly negative above $\sigma_{c}$, but abruptly flips and becomes more positive for $\sigma \geq 2000$Pa (Fig. \ref{Figure1}C, open squares). On increasing $\phi$, N$_{1}$ becomes positive beyond $\sigma_c$ and suggests that the increase in $\eta$ is due to the formation of frictional contacts (Figure 1B, open pentagons and open stars). 

As in previous studies, $\eta_N(\phi)$ and $\eta_{ST}(\phi)$ for our colloidal rod suspensions follow the relations $\eta_N(\phi) = (1-\phi/\phi_{o})^{-2}$ and $\eta_{ST}(\phi) = (1-\phi/\phi_{m})^{-2}$, respectively \cite{Krieger}. Since frictional contacts proliferate only beyond $\sigma_c$, $\phi_{o}$ and $\phi_m$ correspond to viscosity divergences for frictionless and frictional particles, respectively (Fig. \ref{Figure1}D). The presence of these distinct diverging branches in viscosity in shear-thickening suspensions is a key assumption of the recent friction-based Wyart-Cates (WC) model \cite{Cates_PRL} which predicts 

\begin{equation}
\eta(\sigma,\phi) =(1 - \phi/\phi_c(\sigma))^{-2}
\label{eqn1}
\end{equation}

where $\phi_{c}(\sigma)$ = $f\phi_{m} + (1 - f)\phi_{o}$ and $f$ - the fraction of frictional contacts. Further, $f(\sigma,\phi) = f_{max}(\phi) e^{-\sigma^{*}/\sigma}$ where $\sigma^{*}$ sets the threshold stress for particles to form contacts and both $f$ and $\sigma^{*}$ are adjustable parameters in the WC model. Our flow curves can be fit reasonably well to eqn. \ref{eqn1} upto a $\phi = 0.30$ (solid lines in Fig. \ref{Figure1}B) and allowed us to determine $\sigma^*$ and $f_{max}$ at various $\phi$s (Fig. \ref{Figure1}E). Although the spread in $\sigma^*$ is slightly larger than those observed in suspensions of colloidal spheres \cite{Hudson_PRL}, the behavior of $f_{max}$ with $\phi$ is similar. Furthermore, like in \cite{Hudson_PRL}, even to yield a $\beta$ of only 0.15 ($\phi = 0.22$) $f_{max}$ is already as large as 0.85. This suggests that even during CST particle contacts are stabilized by frictional forces, but these contacts remain confined within the hydroclusters which cannot yet percolate to result in a positive N$_{1}$ like in DST. At higher $\phi$s, we find substantial deviations between the model fits and our flow curves and for $\phi = 0.38$, $\eta_{ST}$ is severely underestimated even for $f = 1$. This is not entirely surprising given that the WC model is for isotropic particles. Thus, the development of orientational order under shear in anisotropic particles and the possibility that $f$ might itself depends on the relative orientation of neighboring particles remain unaccounted for within the model.

\subsection*{Unsteady Flow During DST}
Even while the WC model anticipates shear jamming for $\phi\geq\phi_m$, our experiments find that the suspension at $\phi = 0.45 >\phi_m$ flows (stars in Fig. \ref{Figure1}B). Fluids that show DST often have a S-shaped flow curve \cite{Bonn_PRE} with an intermediate region where $d \sigma/ {d\dot{\gamma}} < 0$ (See Supplementary Fig. S2). Since homogeneous flow is impossible in these regions, the system must shear band. In many complex fluids, the canonical example being shear-thinning wormlike micellar gels \cite{Bandyopadhyay_PRL2000,Ganapathy_PRL2006}, the bands themselves can become unstable and can exhibit \textit{rheochaos} - time-dependent flows at fixed $\sigma$ or $\dot{\gamma}$ at Reynold's number $Re << 1$ \cite{Olmsted_RheoActa}. In dense non-Brownian suspensions where steady-state bands are not expected, the suspension flows for $\phi\geq\phi_m$ but this flow is unsteady and exhibits rheochaos in the DST regime \cite{Poon_JoR,Manneville,Heussinger_PRE}. More interestingly, even in Brownian suspensions, experiments have uncovered similar instabilities\cite{Lootens_PRL,vikram_PNAS}. Such complex flow response is not captured by the WC theory which is essentially a mean-field model. Motivated by these findings, we set out to ascertain if the observed bulk flow at $\phi>\phi_m$ (stars in Fig. \ref{Figure1}B) is accompanied by similar instabilities.

\begin{figure*}[htbp]
\centering
\includegraphics[width=0.8\textwidth]{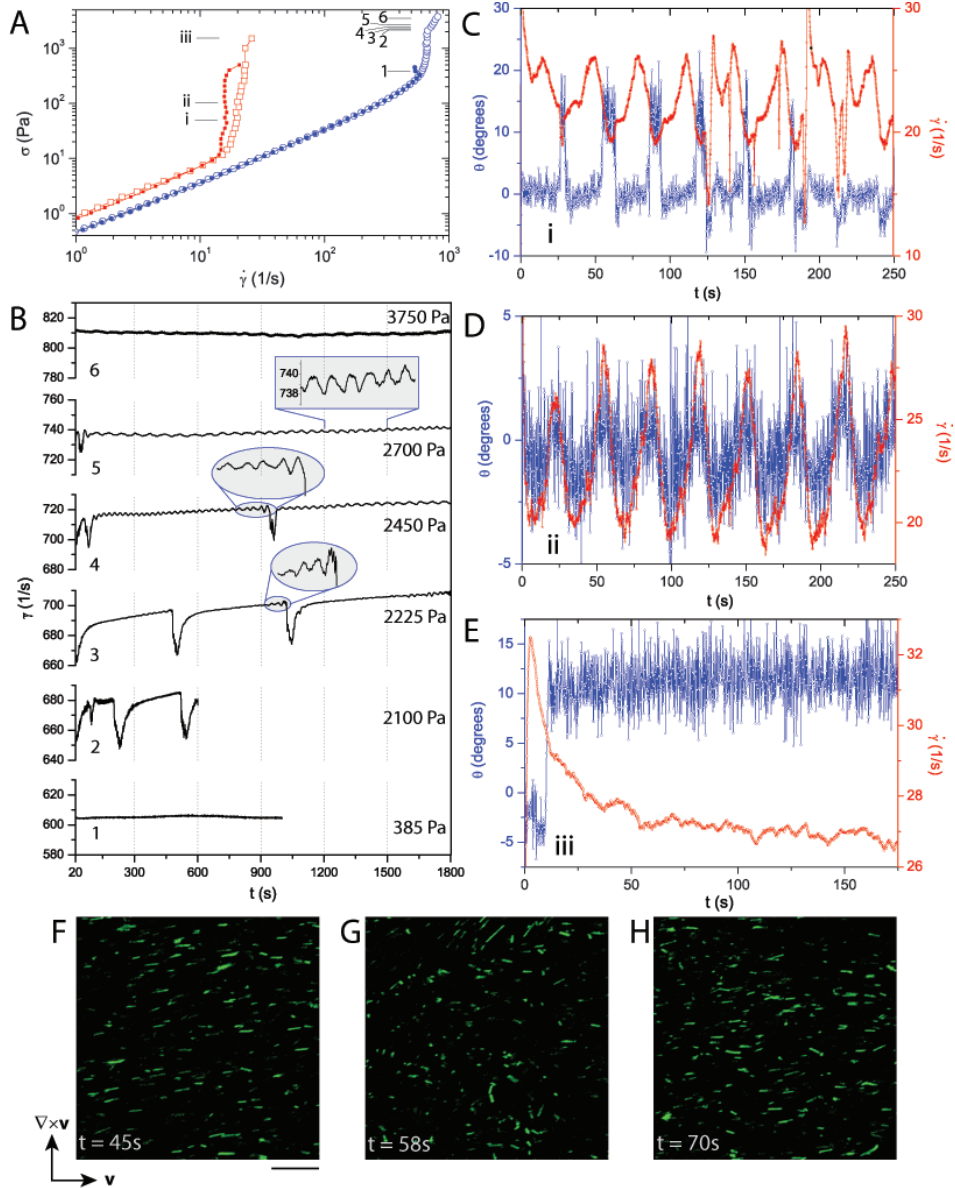}
\caption{(A) Flow curves for $\phi = 0.36$ (blue squares) and $\phi = 0.4$ (red circles) samples. Forward and reverse stress sweeps are shown by hollow and solid symbols, respectively. The stress values at which shear-rate relaxation measurements were carried out are indicated on the flow curves. (B) Time-dependence of shear-rate, $\dot{\gamma}$, at various $\sigma$s for the $\phi = 0.36$ sample. The panel numbers in (B) correspond to those indicated in flow curve ((blue squares) in (A). (C-E) Time-dependence of $\dot{\gamma}$ (red symbols) and the angle $\theta$ between $S$ and $\textbf{v}$ (blue symbols) at various $\sigma$s for the $\phi = 0.4$ sample. (C) $\sigma = 50$ Pa. (D) $\sigma = 100$ Pa. (E) $\sigma = 1500$ Pa. (C), (D) and (E) correspond to points labeled (i), (ii) and (iii) on the flow curve (red circles). (F-H) snapshots of tracer rod orientations over a window spanning a oscillation in $\dot{\gamma}$ for $\sigma = 50$ Pa.} 
\label{Figure2}
\end{figure*}

Accordingly, we carried out shear rate relaxation measurements at various imposed stresses. Figure. \ref{Figure2}A shows forward (hollow symbols) and return (solid symbols) flow curves for $\phi = 0.36 < \phi_m$ (circles) and for $\phi = 0.4 \approx \phi_m$ samples, respectively. Figure \ref{Figure2}B summarizes the results of shear rate relaxation measurements at a few representative stresses for the $\phi = 0.36$ sample. When $\sigma = 385$ Pa $\approx\sigma_c$, where the flow curve is single-valued, $\dot{\gamma}$ shows no time-dependence (Panel 1 in Fig. \ref{Figure2}B). While for regions of the flow curve where the stress is multi-valued (Panels 2-5), $\dot{\gamma}$ shows rich temporal dynamics. At $\sigma = 2100$ Pa, $\dot{\gamma}$ first increases and then saturates before rapidly dropping close to its original value. This behaviour is cyclic. On increasing $\sigma$ (Panels 3 - 5), not only does the cycle duration increase substantially, we also see small amplitude oscillations preceding the drop. The amplitude of these smaller oscillations grows while the time-period shrinks before the drop. Such complex oscillations in the shear-thickening regime have not been observed before and appear to be unique to suspensions of colloidal rods. At even larger $\sigma$s (Panel 6 in Fig. \ref{Figure2}B), where the flow curve is once again single-valued, the time-dependence entirely vanishes.

To gain microscopic insights into these observations, we tracked the dynamics of rods during  shear rate relaxation measurements using a confocal-rheometer \cite{Gokhale_PNAS,Sudeep_RSI}. Owing to the very high particle velocities in the ST regime for the $\phi = 0.36$ sample, particles could not be tracked and the $\phi$ was increased to 0.4. The sample was also doped with a small amount of fluorescently labeled rods \cite{Kuijk_2} to help track their local orientation under shear. At this $\phi$, $\sigma_c \approx 10$ Pa and the $\dot{\gamma}$ values in the DST window are substantially smaller. We also observed hysteresis during the forward (Fig. \ref{Figure2}A open squares) and reverse (Fig. \ref{Figure2}A closed squares) stress sweeps. Figure \ref{Figure2}C shows a portion of the $\dot{\gamma}$ time series (red symbols) for $\sigma$ = 50 Pa. We see oscillations with a nearly well-defined time period of roughly 30s. Simultaneously, we measured the average orientation of the rods and calculated the angle $\theta$ between the global nematic director, $\textbf{S}$, and the flow direction $\bf{v}$ (blue symbols in Fig. \ref{Figure2}C). At $\sigma = 50$ Pa (Fig. \ref{Figure2}C), we observed oscillations in rod orientation between a flow aligned state (Fig. \ref{Figure2}F and H) and a considerably more disordered one (See Supplementary Movie 1). A representative snapshot corresponding to this state (Fig. \ref{Figure2}G) shows a substantial number of rods oriented along the $\nabla v$ direction (seen as circles since the rod orientation is perpendicular to imaging plane) and many rods also tilted towards the vorticity direction. Most remarkably, abrupt increases in $\theta$, i.e. the disruptions in flow alignment, are highly correlated with rapid drops in $\dot{\gamma}$ and therefore correspond to a higher than average viscosity. At $\sigma = 100$ Pa, we once again see that the oscillations in $\dot{\gamma}$ and $\textbf{S}$ are in phase, but here $\theta$ changes gradually (Fig. \ref{Figure2}D and Supplementary Movie 2). For $\sigma = 1500$ Pa, where the flow curve is single-valued, the amplitude of oscillations in $\dot{\gamma}$ is considerably smaller, like in the lower $\phi$ sample, but the dynamics in $\dot{\gamma}$ and $\textbf{S}$ appears noisy (Fig. \ref{Figure2}E)) (Supplementary Movie 3). Our particle-resolved experiments have thus helped unveil, for the first time, the existence of oscillations in particle orientational order and its coupling to bulk rheology during DST and stands in contrast to findings from previous Rheo-SANS measurements \cite{Wagner_2}.

\subsection*{Tuning Inter-particle Friction}
Clearly both inter-particle friction as evidenced by $N_1$ (Fig. \ref{Figure1}C) and a transition to a disordered state (Fig. \ref{Figure2}E) seem to be necessary for observing shear-thickening in colloidal rod suspensions. To isolate the role of frictional contacts during shear-thickening, we designed colloidal rods with tribological properties that could be altered \textit{in situ}. Our approach leaned on recent developments that enable tuning the friction coefficient, $\mu$, between surfaces by grafting them with polymer brushes that undergo a temperature/$pH$-induced coil-globule transition \cite{Chang_Lang,Nordgren_Nanolett,Wu_MRC,Yu_SM}. We restrict our attention to situations where temperature is the external stimuli. On approaching the transition temperature $T_{LCST}$ from below, the enhancement in areal polymer density due to brush collapse creates increasingly stiffer surface asperities and $\mu$ is found to increase by almost an order-of-magnitude. For the specific case of poly N-isopropylyacryl amide (PNIPAm) brushes ($T_{LCST} = 32^\circ$C), $\mu$ is in fact non-monotonic with a maximum at $T_{LCST}$ \cite{Yu_SM}. Beyond $T_{LCST}$, where the brush is fully collapsed, the surfaces adhere due to hydrophobic interactions.

\begin{figure*}[htbp]
\centering
\includegraphics[width=0.65\textwidth]{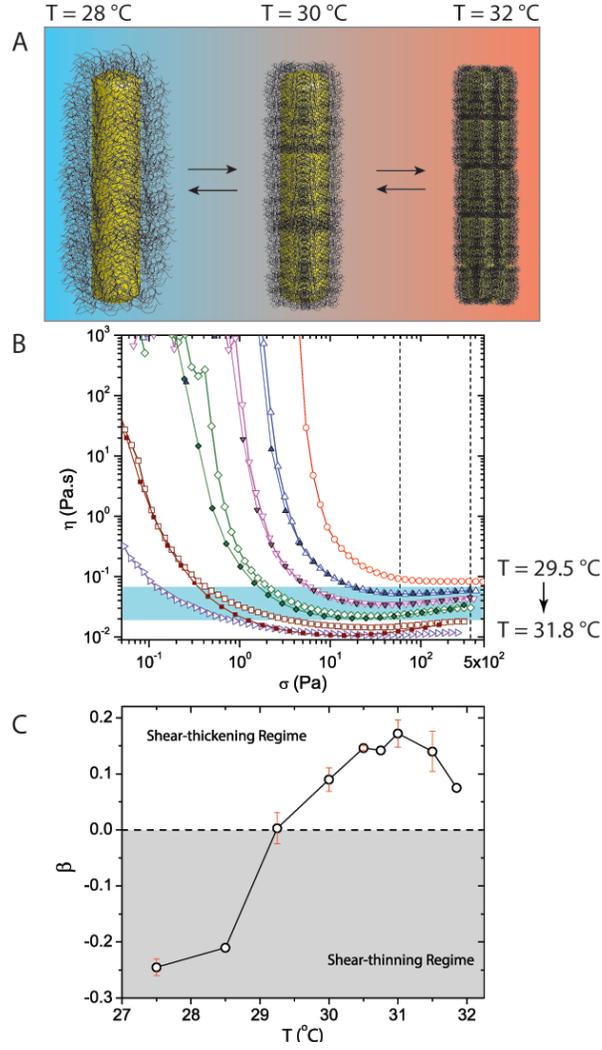}
\caption{(A) Schematic of PNIPAm-grafted rods at temperatures below $T_{LCST}$. (B) Flow curves for PNIPAm-grafted rods at various temperatures below $T_{LCST}$. Hollow symbols represent forward stress sweeps and solid symbols reverse sweeps. $29.5^\circ$C (red circles), $30^\circ$C (blue up-triangles), $30.5^\circ$C (violet down-triangles), $31^\circ$C (green diamonds), $31.5^\circ$C (brown squares) and $31.8^\circ$C (blue right-triangles). (C) $\beta$ versus $T$. The error bars are from multiple experimental realizations.} 
\label{Figure3}
\end{figure*}

We grafted PNIPAm brushes on colloidal silica rods using established routes \cite{Guo} (see Fig. \ref{Figure3}A, Methods and Supplementary text and Fig. S3). Owing to the relatively low yield of particles from our synthesis, all flow measurements were performed in a parallel-plate geometry (gap = $70\mu$m). The flow curves for $T>T_{LCST}$ showed a large hysteresis between the forward and the reverse stress sweeps due to particle aggregation (see Supplementary Fig. S4) and we turned our focus to a relatively narrow temperature window just below $T_{LCST}$ (Fig. \ref{Figure3}B). At low temperatures, prior to undergoing rapid shear-thinning, the suspension also displayed hallmark yield stress fluid behavior and the magnitude of the yield stress decreased on approaching $T_{LCST}$. Most remarkably, while at $T = 29.25^\circ$C shear-thinning culminates in a viscosity plateau that spans almost a decade in $\sigma$ (dashed vertical lines in Fig. \ref{Figure3}B), at higher temperatures we observed CST over the same window of stresses. Perhaps the most surprising finding is that even while the effective volume fraction decreases monotonically on approaching $T_{LCST}$ due to brush collapse, $\beta$ is non-monotonic with a maximum at $T=31^\circ$C (Fig. \ref{Figure3}C).

We now present arguments that suggest that the increase in $\beta$ arises from a concomitant increase in $\mu$. We note that in our experiments due to collapse of the polymer brush on approaching $T_{LCST}$, the effective rod aspect ratio ($\alpha$) increases albeit much less rapidly then the decrease in $\phi$. Increasing $\alpha$ is known to enhance shear-thickening. However, in the rather narrow temperature window between $T = 29.5^\circ$C and $T = 31^\circ$C, where $\beta \geq 0$ and also increases, we expect only a minor change in $\alpha$ and consequently in the dynamical behavior of the rods. This suggests that when frictional contacts cannot form, an order-disorder transition alone \cite{Hoffman} cannot result in shear-thickening. An alternate scenario where $\beta$ can increase is when a large yield stress and/or a shear-thinning stress partially, if not completely, obscures the samples' shear-thickening response \cite{brown_NatMat}. On lowering $\phi$ (increasing temperature here) as more of the shear-thickening regime becomes accessible, $\beta$ can increase. In such cases, however, the high-shear viscosity plateau is practically nonexistent \cite{brown_NatMat}. This is not the case here at any $T\geq 29.5^\circ$C and cannot account for the growth in $\beta$ either. The eventual decrease in $\beta$ can also be easily rationalized. For $T \leq 31^\circ$C forward (hollow symbols) and reverse (solid symbols) stress sweeps coincide, while for $31^\circ\text{C} < T < T_{LCST} $ we observed some hysteresis develop at large stresses (brown symbols in Fig. \ref{Figure3}B) indicating the onset of weak adhesion between particles. Attractive interactions are known to suppress shear-thickening \cite{brown_RPP,Cates_PRL} and this coupled with the decrease in $\phi$ with $T$ should result in a decrease in $\beta$. We emphasize that for $31^\circ\text{C} < T < T_{LCST}$ a second forward stress sweep traced the first indicating the absence of any irreversible particle aggregates (see Supplementary Fig. S5).

\section*{Conclusions}
The collective picture which emerges from our study unambiguously shows that shear-thickening even in suspensions of anisotropic particles has its origins in particles forming a frictional contact network. Not surprisingly, existing friction-based models developed for isotropic particle suspensions do not capture the high $\phi$ flow behavior. Most importantly, we not only find bulk flow in the shear-thickening regime to be unsteady but also provide direct evidence that such  behavior emerges from the strong coupling between the imposed flow and particle orientational order. This should motivate new friction-based theories where particle shape and their associated flow instabilities are explicitly considered. Furthermore, by synthesizing colloidal rods with grafted thermo-responsive polymer brushes, we have taken the first steps towards achieving \textit{in-situ} tunability of $\mu$ which according to recent studies \cite{Fernandez_PRL,Denn_PRL,Cates_PRL,Lin_PRL,Hudson_PRL,Hsu_PNAS} is a key parameter that governs suspension rheology. By altering the thickness of the grafted brush, it should in principle be possible to control the extent of shear-thickening \cite{Zhou_SM}. Further, for an appropriate choice of the polymer, the external stimulus can also be \textit{pH} or light \cite{Niskanen} and incorporation of a suitable copolymer during the grafting step can alter both the location and the width of stimulus response window \cite{Lue}. We believe our strategy presents a multitude of ways to actively tune suspension rheology and is a powerful alternate route to existing methods \cite{Itai_PNAS}. As such our approach should enable active control of a fluids' flow response in demanding technologies that range from microfluidics to body armor and robotic actuators. From merely being an intriguing response of a suspension to flow, our study shows that shear-thickening now lies squarely in the field of materials research.

\section*{Methods}
Silica colloidal rods were synthesized following the procedure developed in \cite{Kuijk_JACS}. Briefly, 70 g of polyvinyl pyrrolidone (PVP) was dissolved in 700 ml of pentanol in a 1000 ml round bottom flask. Then, 70 ml of ethanol followed by 19.6 ml of Milli-q water (Resistivity 18.2 M$\Omega$) and 4.57 ml of sodium citrate dihydrate (0.18M) were added. The suspension was sonicated for 10 mins and vortexed to make sure all the ingredients were thoroughly mixed. Subsequently, we added 15.75 ml of fresh ammonia (28.5 wt\% in water) and 7 ml of tetraethylyortho Silicate (TEOS) and the contents were mixed well. The reaction was left to proceed at 40 $^{o}$C for a duartion of 14 hrs. The rods were cleaned thoroughly in ethanol first and then in water by first centrifuing and then resuspending them in fresh solvent. The suspension was left standing to suspend the clumps and larger rods, which are few in number, and the supernatant which contained rods of a well-defined size and aspect ratio was used in the experiments. To make the rods fluorescent for confocal imaging studies we followed the protocol given in \cite{Kuijk_2}. The procedure was identical to the one mentioned above, expect in the last step we added a mixture of TEOS and aminopropylytriethoxy silane (APS). After cleaning of the rods, we added 25 mg of fluorescein isothiocyanate (FITC) to render the rods fluorescent. Subsequently, another silica layer was grown on the rods to make them stable. To achieve this, the rods were dispersed in 50 ml of ethanol and then 6 ml of ammonium hydroxide 5ml of Milli-q water and finally 0.5 ml of TEOS was added. The reaction was left to proceed for 12 hrs under constant stirring. The rods were then cleaned as before.  The following protocol was followed to synthesize stimuli-responsive core-shell silica rods: the rods were first functionalised with 3-Trimethoxysilylpropylmethacrylate (TPM) \cite{Guo}. The functionalised silica rods were used for precipitation polymerisation of poly N-isopropyl acrylamide. Here, 0.3425 g of PNIPAm and 0.0085 g of crosslinker N,N-methylenebisacrylamide (BIS) was added to 50 ml of 10wt\% rods in water. Nitrogen was bubbled through the solution for 30 mins to remove dissolved oxygen and then 0.0025 g of initiator potassium persulphate (KPS) was added. The reaction was allowed to proceed for 4 hrs at 70$^\circ$C in nitrogen atmosphere. The rods were then cleaned in water to remove the unreacted monomer. 

Bulk rheological measurements were carried out on stress-controlled MCR 301-WESP and MCR 702, rheometers (AntonPaar, Austria) using a cone-plate (cone angle 2$^\circ$, diameter 25 mm) or parallel-plate (diameter 25 mm) geometry. For imaging the fluorescently labeled tracer rods in a shear-flow, the MCR 301-WESP was mounted on an inverted confocal microscope Leica SP5 (Leica, Germany). The tracer rod to bare rods ratio was roughly 1:100. These experiments were done in a cone-plate geometry (cone angle 2$^\circ$, diameter 25 mm) and a 60X objective was used to image the rods. The imaged field of view was 140 $\mu$m $\times$ 140 $\mu$m and the images were captured at 14 frames per second. Subsequent image analysis was done using ImageJ software.

\section{Acknowledgments}
AKS thanks Dept. of Sci. and Tech. (DST), Govt. of India for a Year of Science Fellowship. RG thanks ICMS \& SAMat, JNCASR for support.

\newpage

\newpage  

\newpage

\end{document}